\pdfoutput=1

\documentclass{article}

\usepackage{PRIMEarxiv}

\usepackage[utf8]{inputenc} 
\usepackage[T1]{fontenc}    
\usepackage{hyperref}       
\usepackage{url}            
\usepackage{booktabs}       
\usepackage{amsfonts}       
\usepackage{nicefrac}       
\usepackage{microtype}      
\usepackage{lipsum}
\usepackage{fancyhdr}       
\usepackage{graphicx}       
\usepackage{multirow}
\usepackage{array}
\usepackage{algorithm}
\usepackage{algorithmicx}
\usepackage{algpseudocode}
\usepackage{amsmath}
\usepackage{amssymb}
\usepackage{mathtools}
\usepackage{enumerate}

\graphicspath{{/}}     

\title{GPU‑Accelerated Interpretable Generalization for Rapid Cyberattack Detection and Forensics
}

\author{
  Shu-Ting Huang \\
  Bachelor Program of Big Data Applications in Business \\
  National Pingtung University \\
  \texttt{s9902550173@gmail.com} \\
   \And
  Wen-Cheng Chung \\
  Bachelor Program of Artificial Intelligence \\
  National Yunlin University of Science and Technology \\
  \texttt{sukoyachung@gmail.com} \\
  \AND
  Hao-Ting Pai \thanks{Corresponding author} \\
  Department of Big Data Business Analytics \\
  National Pingtung University \\
  \texttt{haotingpai@gmail.com} \\
}

\begin{document}
\maketitle

\begin{abstract}
The Interpretable Generalization (IG) mechanism recently published in IEEE Transactions on Information Forensics and Security delivers state‑of‑the‑art, evidence‑based intrusion detection by discovering coherent normal and attack patterns through exhaustive intersect‑and‑subset operations—yet its cubic‑time complexity and large intermediate bitsets render full‑scale datasets impractical on CPUs. We present IG‑GPU, a PyTorch re‑architecture that offloads all pairwise intersections and subset evaluations to commodity GPUs. Implemented on a single NVIDIA RTX 4070 Ti, in the 15k‑record NSL‑KDD dataset, IG‑GPU shows a 116× speed‑up over the multi‑core CPU implementation of IG. In the full size of NSL‑KDD (148k-record), given small training data (e.g., 10\%-90\% train-test split), IG‑GPU runs in 18 minutes with Recall 0.957, Precision 0.973, and AUC 0.961, whereas IG required down‑sampling to 15k-records to avoid memory exhaustion and obtained Recall 0.935, Precision 0.942, and AUC 0.940. The results confirm that IG-GPU is robust across scales and could provide millisecond‑level per‑flow inference once patterns are learned. IG‑GPU thus bridges the gap between rigorous interpretability and real‑time cyber‑defense, offering a portable foundation for future work on hardware‑aware scheduling, multi‑GPU sharding, and dataset‑specific sparsity optimizations.
\end{abstract}

\keywords{Intrusion Detection Systems \and Inherently Interpretable Models \and High Computational Efficiency}

\section{Introduction}
Computational efficiency is critical to explainable AI‑based intrusion detection in the IoT \cite{Pham2023ExplainableIDS}. Millions of heterogeneous devices—restricted to milliwatt‑level power and kilobyte‑level memory—must run ML/DL models in real time on multimodal traffic without raising false alarms. Cloud offloading lessens the load but adds latency and energy costs, and cutting‑edge neural networks remain too heavy for latency‑sensitive realms like vehicular IoT. Research is therefore pivoting to fog and edge‑centric architectures that move heavy analytics nearer the data, pursuing a balanced trio of speed, scalability, and interpretability for trustworthy cyber defense.

Interpretable Generalization (IG) \cite{Pai2024IGM} breaks the prevailing trade‑off between speed and interpretability. By discovering coherent patterns—feature subsets exclusive to either normal or malicious flows—during training, IG unifies detection and explanation. In publicly available datasets, IG consistently delivers high accuracy regardless of the size of the training data, while providing line-by-line intrusion paths suitable for forensic audits. The exhaustive intersect‑and‑subset search that powers this insight, however, scales cubically with the number of instances and generates gigabyte‑scale bitsets, limiting practical use to down‑sampled datasets and off‑line analysis.

We introduce IG‑GPU, a PyTorch re‑architecture that off‑loads pairwise intersections, subset filtering, and score accumulation to commodity graphics processors. Bit‑packed int64 tensors, dynamic batching, and device‑agnostic kernels enable a single NVIDIA RTX 4070 Ti to process the NSL‑KDD within tens of seconds—around 116× wall‑clock acceleration over a multi‑core CPU—while preserving IG’s outstanding Precision, Recall, and AUC across both full‑scale and 10\% subset workloads. By fusing intrinsic interpretability with near–real‑time throughput, IG‑GPU dissolves the longstanding trade‑off between transparency and efficiency in IDS.

\section{Related Work}
The study \cite{Pham2023ExplainableIDS} in IEEE Communications Surveys \& Tutorials offers an authoritative panorama of explainable IDS: most detectors first optimize for accuracy with opaque learners and afterwards bolt on interpretability through post‑hoc toolkits such as SHAP and LIME. Despite methodological breadth, three liabilities press for intrinsically interpretable and hardware‑efficient solutions. Fidelity: surrogate explainers can stray from the black box when inputs drift or when  adversaries attack the explanation itself. Cost: producing saliency maps or feature‑importance scores can raise latency and memory 10-100×, blowing past 10 Gbps and edge‑gateway limits. Scalability: datasets like the 147k‑record NSL‑KDD already tax CPUs; adding an explainer only widens the gap.

These findings sharpen the case for algorithms whose interpretability is intrinsic and whose computation maps naturally onto parallel hardware. Interpretable Generalization satisfies the first criterion by deriving coherent patterns that serve simultaneously as decision rules and human‑readable evidence, yet its cubic intersect‑and‑subset search threatens real‑time performance. The IG‑GPU architecture presented in this work targets the second criterion, demonstrating that intrinsic explainability and commodity‑level speed can indeed coexist.

\section{Methodology}
\subsection{Interpretable Generalization Algorithm}
Consider a discretized feature universe $v = \{v_1, v_2, \ldots, v_d\}$. Each network flow is encoded as a bitset $x \subseteq V$: numeric attributes are z-scored and rounded to \textit{p} decimals, categorical values are taken verbatim, and every value is concatenated with its column index to form a unique token. A coherent pattern is a token subset that appears only in one class. IG discovers such patterns in two passes.

First, within each class $c \in \{+, -\}$ equation (1) enumerates every intersection, retaining both pairwise overlaps and full instance signatures. Each candidate $b \in B^c$ is assigned as equation (2), where $f_c(b)$ counts supporting instances and $|b|$ is the bit length, so richer patterns weigh more heavily.
\begin{equation}
B^c = \{x \cap x' \mid x, x' \in X^c,\, x \leq x'\} \cup X^c
\end{equation}
\begin{equation}
S_c(b) = f_c(b) \, |b|^2
\end{equation}

Second, in equation (3) IG expels every candidate that ever appears inside the opposite class, yielding the coherent dictionaries. At test time a flow \textit{x} receives two aggregate evidence by equation (4), where $\mathbb{I}[\![\,\cdot\,]\!]$ is the indicator function.
\begin{equation}
P^c = \{b \in B^c : b \not\in x \text{ for all } x \in X^{\overline{c}}\}
\end{equation}
\begin{equation}
A(x) = \sum_{b \in P^+} S_+(b) \, \mathbb{I}[b \subseteq x], \quad
N(x) = \sum_{b \in P^-} S_-(b) \, \mathbb{I}[b \subseteq x]
\end{equation}

Next, a classifier is generated by the following regulations. Regulation 1 (primary comparison). If $A(x) \geq N(x)$, label $x$ attack; otherwise normal. Regulation 2 (zero-evidence default). If $A(x) = 0 \;\&\; N(x) = 0$, override Regulation 1 and label $x$ attack. Regulation 3 (outlier within “normal” evidence). If $N(x) < \mu_N - (r \times \sigma_N)$, label $x$ attack even when Regulation 1 predicts normal; else keep the label from Regulation 1.

\subsection{Data Layout}
Throughput on contemporary accelerators is limited less by floating-point capacity than by the rate at which data can be moved to, from and within device memory. After feature discretisation each network flow is represented as a Boolean vector $b_i \in \{0,1\}^L
$. To align with the native 64 bit datapaths of CUDA, using equation (5), 64 successive bits are aggregated into a single machine word, yielding the packed row.
\begin{equation}
B_i = (b_i^{(0)}, \ldots, b_i^{(K-1)}) \in \text{uint64}^K,\; K = \lceil L / 64 \rceil
\end{equation}
This transformation divides the footprint of every row by 64, converts bitwise AND into a warp synchronous single instruction operation, and ensures that global-memory transactions are fully coalesced. Rows are then stacked by class into two contiguous tensors, $\text{Rows}A \in \mathbb{Z}_2^{n{_A} \times K}$ for anomalies and $\text{Rows}N \in \mathbb{Z}_2^{n{_A} \times K}$ for normals. Contiguity removes pointer chasing inside kernels and enables trivial pointer arithmetic for batched slicing. Because PyTorch exposes int64 but not uint64 tensors on both CUDA and Metal, each word is stored as two’s-complement int64; the truth table of bitwise AND is unchanged by this re interpretation.

\subsection{Formal Operations}
IG’s three logical stages—pairwise pattern enumeration, cross-class subset rejection and score accumulation—can all be framed as tensor-algebraic primitives that map naturally onto GPU hardware. \textit{Pairwise intersection} within class $C$ is the blocked Hadamard product $(\hat{\mathbf{R}}_c \odot \hat{\mathbf{R}}_c^{\top})_{ijk} = \hat{\mathbf{R}}_{C,ik} \land \hat{\mathbf{R}}_{C,jk} \quad (i < j)$, where $\hat{\mathbf{R}}_c \in \text{uint64}^{n_c \times K}$ is the packed row tensor. In practice we fix the left index $i$ and broadcast-AND it over a sliding window of $B$=\text{PAIR\_BATCH} peers, so that the entire $O(n_c^2)$ search is executed by $\left\lceil n_c^2 / B \right\rceil$ memory-bound kernels. \textit{Cross-class subset filtering} is realized as a mixed Boolean matrix multiplication expressed in equation (6), followed by $\cdot m_p = \displaystyle\mathop{\bigvee}_{t=1}^{n_{\bar{C}}} \text{S}_{pt}$ , which marks pattern P as \textit{covered} if any opposite-class row is a superset. The inner conjunction and outer disjunction correspond, respectively, to torch.all across the block dimension and torch.any across rows; both execute as warp-level vote instructions.
\begin{equation}
\text{S}_{pt} = \displaystyle\mathop{\bigwedge}_{k=1}^{K} \left[ \left( \hat{\mathbf{P}}_{pk} \land \hat{\mathbf{R}}_{\bar{C},tk} \right) = \hat{\mathbf{P}}_{pk} \right]
\end{equation}
\textit{Inference} reduces to a masked tensor-vector contraction. Let $\hat{\mathbf{T}} \in \text{uint64}^{n_{test} \times K}$ be the packed test matrix and $s \in \mathbb{Z}^{|P_{\text{pure}}|}$ the vector of pattern scores. Boolean coverage $\text{Z}_{pt} = \displaystyle\mathop{\bigwedge}_{k=1}^{K} \left[ \left( \hat{\mathbf{P}}_{pk} \land \hat{\mathbf{T}}_{tk} \right) = \hat{\mathbf{P}}_{pk} \right]$ is multiplied by s and reduced in one fused kernel, producing abnormal and normal score vectors that reside entirely in device memory until the final copy to host.

\begin{figure}
  \centering
  \includegraphics[width=8.47cm]{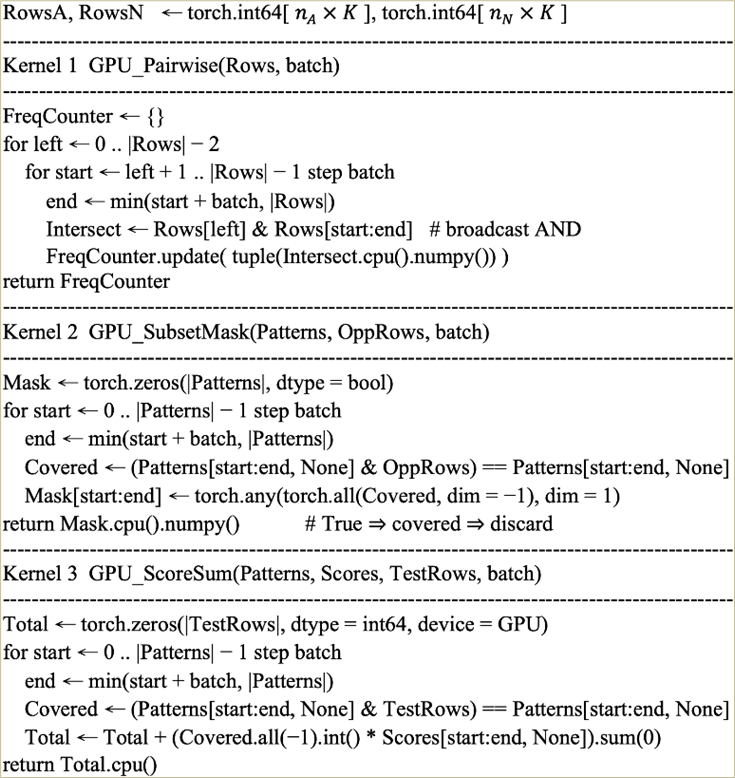}
  \caption{GPU-accelerated IG kernels}
  \label{fig:fig1}
\end{figure}

Fig. \ref{fig:fig1} presents IG-GPU’s kernel pipeline. \textit{Kernel 1} materializes up to \textbf{batch} intersection vectors per invocation (where batch is a tunable parameter rather than a fixed value) and transfers them to the host in a single burst, minimizing PCIe traffic. \textit{Kernel 2} merges a three-dimensional Boolean tensor down to one dimension without leaving VRAM. \textit{Kernel 3} fuses coverage testing, integer weighting and reduction into a single pass whose roofline position is firmly memory-bound.

\section{Experiments}
To attribute every reported speed‑up exclusively to the GPU kernel, we follow the subset protocol of IG \cite{Pai2024IGM}. Table 1 therefore benchmarks three curated data sets. For NSL‑KDD we adopt the canonical 15,000‑record slice—10,000 consecutive rows from KDDTrain+ followed by 5,000 from KDDTest+—yielding 7,439 normal and 7,561 anomalous flows across 41 features. For UNSW‑NB15 we replicate the 14,174‑record construction in \cite{Pai2024IGM}, comprising the first 10,000 normal flows encountered in the four CSV shards and 4,174 anomalies gathered by iterating through the nine attack categories until each supplies up to 500 instances, giving 47 features per record. The UKM‑IDS20 corpus is retained intact, totaling 12,887 flows (8,909 normal, 3,978 anomalous, 46 features). Each subset then traverses a common five‑stage pipeline—conversion→preservation→z‑score discretization→ column re‑encoding→anti‑contradiction filtering—with no additional sampling, weighting, or feature elimination. For scale‑out validation, Table 2 reports accuracy on the complete 148,517‑record NSL‑KDD data set.

All benchmarks were run on a workstation featuring an AMD Ryzen 9 5950X (16 cores / 32 threads, 128 GB DDR4‑2,666) for the CPU baseline and a single NVIDIA RTX 4070 Ti (7,680 CUDA cores, 12 GB GDDR6X) for the IG‑GPU configuration. The PyTorch 2.3 kernels operate on bit‑packed int64 tensors with dynamic batching. Because no device‑specific heuristics are embedded, the implementation is host‑agnostic and scales linearly to higher‑end accelerators.

Table 1 summarizes throughput across nine train/test splits (1|9 through 9|1). IG‑GPU attains speed‑ups ranging from 37× to 430× over the CPU baseline, the latter on UKM‑IDS20, shrinking hours of cubic‑time computation to minutes of wall‑clock time. On average, the accelerations are 116× (NSL‑KDD), 116× (UNSW‑NB15), and 145× (UKM‑IDS20), indicating that the intersect‑and‑subset search maps efficiently to commodity parallel hardware—even when intermediate bitsets exceed on‑board memory. The GPU kernel remains bit‑for‑bit identical to the reference CPU implementation; pattern counts are preserved exactly, ensuring that the observed speed‑ups do not compromise the evidential basis of IG’s decisions.

Executing IG-GPU on the complete NSL-KDD dataset (148,517 records; see Table 2) confirms—and in some metrics surpasses—the performance observed on the 15k sequential subset. With a 1:9 class ratio, the full-scale run attains an accuracy of 0.9666, Recall 0.9573, Precision 0.9732, and AUC 0.9609, completing pattern discovery in 18 min 21 s (1101 s) on a single RTX 4070 Ti. While the down-sampled experiments peak at a precision of 0.978 (30\% training) yet allow recall to drop to 0.844, the full-scale model maintains both precision and recall above 0.957 and preserves F1 and AUC $\geq$ 0.96. Scaling to all 71,463 anomalous flows therefore reduces variance without compromising interpretability or throughput, demonstrating that IG-GPU’s bit-packed tensor kernels convert exhaustive combinatorial evidence into production grade speed and stability—capabilities not yet reported by CPU-bound or post-hoc XAI-based IDSs under comparable workloads. 

These results establish that a single intrusion‑detection system can now deliver intrinsic interpretability, real‑time throughput, and state‑of‑the‑art accuracy. Porting IG’s exhaustive combinatorial kernel to GPUs achieves an accuracy of 0.967, a precision of 0.973, a recall of 0.957, and an AUC of 0.961, while reducing training time by up to 430×. Crucially, the GPU implementation retains byte‑exact pattern evidence, creating two orders of magnitude of additional headroom for multi‑GPU sharding, hardware‑aware scheduling, and sparsity‑driven pruning—without compromising the forensic rigor that motivates IG.

\begin{table}
\tiny
 \caption{CPU vs. GPU Acceleration Time Comparison}
  \centering
  \begin{tabular}{c|rrrr|rrrr|rrrr}
    \toprule
    \multirow{2}{*}{Ratios}     & \multicolumn{4}{c|}{NSL-KDD}     & \multicolumn{4}{c|}{UNSW-NB15}     & \multicolumn{4}{c}{UKM-IDS20}\\
    \cline{2-13}
    & \multicolumn{1}{r}{Patterns} & CPU                  & GPU & Speedup                  & Patterns & CPU                  & GPU & Speedup                  & Patterns & CPU                  & GPU & Speedup \\
    \hline
    1 | 9 & 30942 & 861.46 & 11.13 & 77 & 394713 & 4633.96 & 27.76 & 167 & 19708 & 1282.98 & 4.76 & 270 \\
    2 | 8 & 72850 & 1901.33 & 10.11 & 188 & 1047138 & 13305.89 & 83.02 & 160 & 47541 & 3284.97 & 7.64 & 430 \\
    3 | 7 & 115724 & 2917.78 & 17.98 & 162 & 1855246 & 24363.06 & 160.08 & 152 & 82727 & 5446.81 & 137.12 & 40 \\
    4 | 6 & 162164 & 3984.84 & 30.85 & 129 & 2758367 & 37052.70 & 245.92 & 151 & 120964 & 7537.56 & 102.76 & 73 \\
    5 | 5 & 210448 & 5213.69 & 40.87 & 128 & 3744057 & 49943.85 & 349.92 & 143 & 159847 & 8710.19 & 156.67 & 56 \\
    6 | 4 & 256095 & 6087.76 & 60.90 & 100 & 4813101 & 54962.07 & 477.76 & 115 & 199681 & 9275.38 & 113.43 & 82 \\
    7 | 3 & 277776 & 7214.78 & 75.80 & 95 & 5927126 & 49588.46 & 602.12 & 82 & 238552 & 9404.48 & 66.22 & 142 \\
    8 | 2 & 297033 & 9031.19 & 104.40 & 87 & 7094785 & 34400.88 & 812.44 & 42 & 279776 & 9914.01 & 85.49 & 116 \\
    9 | 1 & 325130 & 9538.21 & 122.96 & 78 & 8294443 & 34754.07 & 938.45 & 37 & 322939 & 9884.24 & 101.17 & 98 \\
    \bottomrule
  \end{tabular}
  \label{tab:table1}
\end{table}

\begin{table}
\tiny
 \caption{Full NSL‑KDD (1:9) Detection Performance}
  \centering
  \begin{tabular}{c|rrrrr}
    \toprule
    \multirow{2}{*}{Ratios}     & \multicolumn{5}{c|}{NSL-KDD (std=0.568)}\\
    \cline{2-6}
    & \multicolumn{1}{r}{Accuracy} & Recall                  & Precision & AUC                  & Time \\
    \hline
    1 | 9 & 0.96662 & 0.95728 & 0.97315 & 0.96093 & 1101\\
    \bottomrule
  \end{tabular}
  \label{tab:table2}
\end{table}

\section{Conclusion}
IG‑GPU elevates Interpretable Generalization from a CPU‑bound prototype to a production‑ready detector. Executing the full 148 k‑record NSL‑KDD corpus on a single US\$ 800 NVIDIA RTX 4070 Ti, it finishes in 18 minutes while attaining Accuracy 0.967, Precision 0.973, Recall 0.957, and AUC 0.961—all with byte‑level evidence retained for forensic auditing. This >100× speed‑up removes the compute‑time and memory constraints that once mandated down‑sampling on multi‑core CPU. Consequently, IG‑GPU shows that strict interpretability and state‑of‑the‑art accuracy can now coexist with real‑time throughput on commodity hardware, transforming exhaustive, evidence‑based intrusion detection from an academic aspiration into a practical foundation of network defense. In contrast, the research \cite{Thompson2020Limits} from MIT shows that deep‑learning accuracy scales by a severe power‑law (exponent $\approx$ 12): halving ImageNet error needs > 4,000× more compute, and even a 10 \% gain costs > 3×. Such runaway demands, plus the overhead of post‑hoc explainers, already cripple real‑time IDS on 10‑Gb links. IG‑GPU avoids this spiral by encoding decisions as GPU‑parallel symbolic rules that double as explanations, achieving transparent, wire‑speed detection on commodity hardware.

\section{Acknowledgments}
This work was supported by the National Science and Technology Council (NSTC), Taiwan, under grant number 114-2221-E-153-009.

\bibliographystyle{unsrt}  
\bibliography{references}

\end{document}